\documentclass[prb,preprint]{revtex4}% Physical Review B
\usepackage{amsmath}
\usepackage{bbm}
\usepackage{epsfig}
\begin{document}
\title{Multiquantum well spin oscillator}
\author{L.L. Bonilla}
\affiliation {Modeling, Simulation and Industrial Mathematics, Universidad
Carlos III de Madrid, 28911 Legan\'es, Spain and Unidad Asociada al Instituto 
de Ciencia de Materiales de Madrid, CSIC.}
\author{R. Escobedo}
\affiliation{Departamento de Matem\'atica Aplicada, Universidad de Cantabria, 39005 
Santander, Spain.}
\author{ M. Carretero}
\affiliation {Modeling, Simulation and Industrial Mathematics, Universidad
Carlos III de Madrid, 28911 Legan\'es, Spain and Unidad Asociada al Instituto 
de Ciencia de Materiales de Madrid, CSIC.}
\author{G. Platero}
\affiliation { Instituto de Ciencia de Materiales, CSIC, Cantoblanco, Madrid, 28049, 
Spain.}
\date{\today}
%%%%%%%%%%%%%%%%%%%%%%%%%%%%%%%%%%%%%%%%%%%%%%%%%%%%%%%%%%%%%%%%%%
%%%%%%%%%%%%
\begin{abstract}
A dc voltage biased II-VI semiconductor multiquantum well structure attached to normal
contacts exhibits self-sustained spin-polarized current oscillations if
one or more of its wells are doped with Mn. Without magnetic impurities,
the only configurations appearing in these structures are stationary.
Analysis and numerical solution of a nonlinear spin transport model
yield the minimal number of wells (four) and the ranges of doping
density and spin splitting needed to find oscillations.
\end{abstract}
%%%%%%%%%%%%%%%%%%%%%%%%%%%%%%%%%%%%%%%%%%%%%%%%%%%%%%%%%%%%%%%%%%
%%%%%%%%%%%%
%\begin{multicols}{2}

\maketitle

Among spintronics challenges, electrical injection of spin polarized current in
semiconductor nanostructures is important due to their
potential applications as spin-based devices \cite{khaetski,mireles}. Different
spin injectors have been proposed, including ferromagnetic contacts or semimagnetic
semiconductor contacts with large g factors that are polarized
by a magnetic field at low temperatures. The efficiency of
ferromagnetic/semiconductor junctions has shown to be very small
due to the large conductivity mismatch between the metal and the semiconductor
\cite{schmidt}. Diluted magnetic semiconductors (DMS) are much more
efficient as spin injectors, as it has been shown for contacts based in Mn
\cite{egues,fiederling,khaetski,david2}.

Much theoretical and experimental work
\cite{bonilla,BGr05,sanchez,inarrea}
is devoted to the analysis of nonlinear transport through
conventional semiconductor superlattices (SLs), in which
the interplay between Coulomb interaction, electron confinement
and dc voltage produces very interesting properties such as
formation of electric field domains (EFDs) and self-sustained current
oscillations (SSCOs). In addition, external ac electric fields
produce additional features in the nonlinear I/V curve such as photo-assisted EFDs and
absolute negative differential resistance in the non-adiabatic
limit \cite{inarrea} or, at low ac frequencies, chaotic current oscillations 
\cite{bulashenko,sanchez1,luo}.

Compared to conventional semiconductor nanostructures, DMS
present an additional degree of freedom: the spin, which plays an
important role in electron dynamics. In particular, II-VI based
semiconductor SLs doped with $Mn^{++}$ ions \cite{crooker}. In
these systems, carrier-ion exchange spin effects dominate the
magneto-transport, producing spin polarized transport and large
magneto-resistance. Exchange interaction between the spin carrier
and Mn ions results in large spin splittings. In fact full spin
polarization has been achieved at magnetic fields of 1 Tesla.
Recently \cite{david}, nonlinear transport through DMS SLs
has been investigated. The interplay between the nonlinearity of
the current--voltage characteristics and the exchange
interaction produces interesting spin dependent features
 \cite{david}: multistability of steady states with different polarization in the magnetic
 wells, time-periodic oscillations of the spin-polarized current and induced
spin polarization in nonmagnetic wells by their magnetic neighbors, among others. The high
sensitivity of these systems to external fields points out to their potential application as
magnetic sensors \cite{david}.

In this letter we analyze nonlinear electron spin dynamics of a n-doped dc voltage biased
semiconductor multiquantum well structure (MQWS) having one or more of its wells doped with Mn.
We show that spin polarized current can be obtained even using normal contacts,
provided one quantum well (QW) is doped with magnetic impurities (Mn).
We analyze under which conditions the system exhibits static EFDs and stationary current
or moving domains and time-dependent oscillatory current. SSCOs appear in nanostructures
with at least four QWs. Moreover, SSCOs may appear or not depending on the spin splitting
$\Delta$ induced by the exchange interaction. From our results we propose how to design a 
device behaving as a spin-polarized current oscillator.

{\it Theoretical  model.}
Our sample configuration consists of an n-doped ZnSe/(Zn,Cd,Mn)Se weakly coupled
MQWS. The spin for the magnetic ion $Mn^{++}$ is S=5/2 and the exchange interaction
between the Mn local moments and the conduction band electrons is ferromagnetic in II-VI
QWs. Using the virtual crystal and mean field approximations, the effect of the exchange
interaction is to make the subband energies spin dependent in those QWs that
contain Mn ions: $E_j^{\pm} =E_j \mp \Delta/2$ where $\Delta=2J_{sd}N_{Mn}S\,
B_S(g\mu_BBS/(k_B T_{\rm eff}))$ for spin $s=\pm 1/2$, and $B$, $J_{sd}$, $N_{Mn}$,
and $T_{\rm eff}$ are the external magnetic field, the exchange integral, the
density of magnetic impurities and an effective temperature which
accounts for Mn interactions, respectively \cite{david,slobo}. We model spin-flip scattering
coming from spin-orbit or hyperfine interaction by means of a phenomenological scattering
time $\tau_{\rm sf}$, which is larger than impurity and phonon
scattering times: $\tau_{ \rm scat}= \hbar/{\gamma} <\tau_{\rm
sf}$. Vertical transport in the MQW is spin-independent
sequential tunneling between adjacent QWs, so that when electrons
tunnel to an excited state they instantaneously relax by phonon
scattering to the ground state with the same spin polarization.
Lastly, electron-electron interaction is considered within the
Hartree mean field approximation. The equations describing our
model generalize those in Ref.~\onlinecite{david} to the case of
finite $T$:
\begin{eqnarray}
&& F_i - F_{i-1} = \frac e{\varepsilon} (n_i^+ + n^-_{i} - N_D), \label{e1}\\
&& e\,\frac{{\rm d}n^\pm_i}{{\rm d}t} =
J^\pm_{i-1\rightarrow i} - J^\pm_{i \rightarrow i+1}
\pm {A(n_i^+,n_i^-,\mu_{i}^+) \over \tau_{\rm sf,i}},
\label{e2}
\end{eqnarray}
$i=1,\dots, N$. $A(n_i^+,n_i^-,\mu_{i}^+)=n_{i}^- - n_{i}^+/\alpha_i$,
with $\alpha_i = 1 + \exp[(E_{1,i}^- - \mu_{i}^+)/\gamma_{\mu})]$.
As $\gamma_{\mu}\to 0$, $A(n_i^+,n_i^-,\mu_{i}^+)$ becomes
$\pm(n^-_{i}-n^+_{i})/\tau_{\rm sf}$ for $\mu_{i}^+ > E_{1,i}^-$ (equivalently,
$\mu_{i}^+ - E_{1,i}^+ > \Delta$), and $\pm n^-_{i}/\tau_{\rm sf}$ otherwise
\cite{david}.

Here $n^+_i$, $n_{i}^-$, $-F_i$ and $\mu_i^\pm$ are the two-dimensional spin-up
and spin-down electron densities, the average electric field and the chemical potential
at the $i$th SL period (which starts at the right end of the ($i-1$)th barrier and finishes
at the right end of the $i$th barrier), respectively.
$E_{j,i}^\pm$ are the spin-dependent subband energies (measured from the
bottom of the $i$th well): $E_{j,i}^\pm = E_{j}\mp \Delta_{i}/2$, with $\Delta_{i}=
\Delta$ or 0, depending on whether the $i$th well contains magnetic impurities.
$N_D$, $\varepsilon$, $l=d+w$, and $-J^\pm_{i\to i+1}$ are the 2D doping density at the
QWs, the average permittivity, $d$ and $w$ are barrier and well width. The tunneling current 
density across the $i$th barrier $J_{i\to i+1}^\pm$
are calculated by the Transfer Hamiltonian method:
\begin{eqnarray}
J_{i\to i+1}^\pm = {e\, v^{\pm} \over l}
\left\{n_i^\pm - \rho
\ln\left[ 1 + e^{{n_{i+1}^\pm \over \rho} - {F_i \over a}} - e^{- {F_i\over a}}
\right] \right\} \label{e5}
\end{eqnarray}
where $i=1,\dots,N-1$, $\rho = m^* k_B T /(2\pi\hbar^2)$, $m^*$ is the effective 
electron mass and $a = k_B T /(el)$. The voltage bias condition is $\sum_{i=0}^N F_i l = V$ 
for the applied voltage $V$. For electrons with spin $\pm 1/2$, $\mu_{i}^\pm$ and $n_{i}^
\pm$ are related by $n_{i}^\pm = (\rho/ N_D)\,\ln \left[ 1 + \exp[(\mu_i^\pm
- E_{1,i}^\pm)/( k_{B}T)]  \right]$. 
Defining $J_{i\rightarrow i+1} = J_{i\rightarrow i+1}^+ +J_{i\rightarrow i+1}^-$,
time-differencing (\ref{e1}) and inserting the result in (\ref{e2}), we obtain
an expression for the total current density $J(t)$ when $dV/dt=0$:
$\varepsilon\, {\rm d}F_i/{\rm d}t + J_{i\rightarrow i+1}
= J(t) = (N+1)^{-1}\sum_{i=0}^N J_{i\rightarrow i+1}.$
As boundary tunnelling currents for $i=0$ and $N$, we use (\ref{e5}) with
$n^\pm_{0}=n_{N+1}^\pm= N_{D}/2$ (identical normal contacts) \cite{david}. Initially, 
we set $F_{i}=V/[l(N+1)]$, $n_{i}^\pm=N_{D}/2$ (normal QWs). The spin-dependent
``forward tunneling velocity'' $v^{\pm}(F_i)$ is a sum of Lorentzians of width $2\gamma$
(the same value for all sub-bands, for simplicity) centered at the resonant field values
$F_{j,i}^\pm= (E_{j,i+1}^\pm-E_{1,i}^\pm)/(el)$ \cite{BGr05}:
\begin{eqnarray}
v^{\pm} (F_i) & = & \sum_{j=1}^2
\frac{ {\hbar^3 l \gamma \over 2\pi^2m^{*2}}\,
{\cal T}_i (E_{1,i}^\pm) }{ (E_{1,i}^\pm - E_{j,i+1}^\pm + eF_il)^2 + (2\gamma)^2}.
\label{e6}
\end{eqnarray}
Here ${\cal T}_i$ is proportional to the dimensionless transmission probability across the
$i$th barrier \cite{BGr05}.

{\it Results.}
We have considered a sample with $d=10$~nm, $w=5$ nm, $\Delta =15$ meV,
$\tau_{\rm sf}= 10^{-9}$ s (normal QW) and $10^{-11}$s (magnetic QW)\cite{aws},
$m^*=0.16 m_{0}$, $\varepsilon=7.1 \varepsilon_{0}$, $T=5$ K,
$E_{1}=15.76$ meV, $E_{2}=61.99$ meV, $\gamma=1$ meV and $\gamma_{\mu}=0.1$ meV.

\vspace{-0.4cm}
\begin{figure}[ht]
\centerline{\hbox{\hspace{-0.3cm}
 \epsfxsize=80mm
\epsfbox{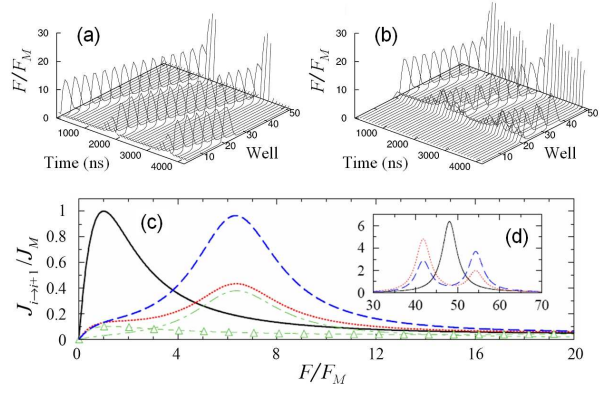} }}
\caption{
Electric field distribution if the magnetic QW is: (a) $i=1$, (b) $i=20$.
(c) Solid line: $J_{i\to i+1}(F)$ for nonmagnetic $i$ and $i+1$.
For magnetic $i$, nonmagnetic $i\pm 1$: $J_{i\to i+1}$ (dotted line), $J^+_{i\to i+1}$ (dot-dashed line),
$J^-_{i\to i+1}$ (triangles), $J_{i-1\to i}$ (dashed line).
(d) same at larger electric fields. Parameter values: $N=50$, $V=0.048$ V, $N_{D}=
10^{10}$cm$^{-2}$, $F_{M}= 0.64$ kV/cm, $J_{M}= 0.409$ A/cm$^2$.}
\label{2}
\end{figure}

There are SSCOs for a variety of configurations, but only if one or more
QWs contain magnetic impurities yielding a sufficiently large spin splitting.
The nonmagnetic MQWS does not exhibit self-oscillations.

Firstly, we have used long SLs ($N=50$), finding that charge dipoles are triggered
at the well containing Mn that is closest to the injector. These dipoles move to the collector 
(near which they may become monopoles if $V$ is large enough), disappear there, and new 
dipoles are triggered, producing SSCOs similar to those observed in III-V semiconductor SLs 
\cite{BGr05}.

Figs.~\ref{2}(a),(b) show that if the only magnetic QW is the $i$th (with $1\leq i<N-3$),
the dipoles are emitted at this well, and their motion is limited to the last $N-i$ QWs.
Why is this?
Fig.~\ref{2}(c) depicts $J_{i\to i+1}(F,N_{D}/2,N_{D}/2)$.
As $E^\pm_{1,i+1}=E_{1}$, $E_{1,i}^\pm = E_{1}\mp\Delta/2$, the $j=1$ term
 in (\ref{e6}) is a Lorentzian centered at $F^\pm_{1,i}=
\pm\Delta/(2el)$. Then $J_{i\to i+1}$
has a peak roughly at $(\Delta^2 + 8\gamma^2)/(2el\Delta)$ (if
$eF_{M}l\ll\Delta/2$), mostly due to $J^+_{i\to i+1}$. The height
of this peak is under half that of $J_{i\to i+1}(F)$ for
nonmagnetic wells ($\mathcal{T}_{i}$ is smaller for $E_{1}^+$ than
for $E_{1}^-$), as depicted in Fig.~\ref{2}(c).
Spin splitting also causes $J_{i\to i+1}$ (for magnetic QW $i$) to display two peaks
at $(E_{2}-E_{1}\pm\Delta/2)/(el)$ instead of one peak at $(E_{2}-E_{1})/(el)$
with their combined strength (for nonmagnetic QW $i$); see Fig.~\ref{2}(d).
Similarly, if QW $i$ is magnetic, $J_{i-1\to i}^\pm$ has peaks at $\mp \Delta/(2el)$
and $(E_{2}-E_{1}\mp\Delta/2)/(el)$, contrary to the shifts in $J_{i\to i+1}^\pm$.

The shifted curves $J_{i-1\to i}$ and $J_{i\to i+1}$ play the role of effective cathode
boundary currents during SSCOs. Clearly, they intersect the current farther away from the
magnetic QW [solid line in Fig.~\ref{2}(c)] on its second, decreasing branch.
The intersection point corresponds to the critical current for triggering a charge
dipole \cite{sanchez,BGr05}.
For Fig.~\ref{2}(b), the boundary current at the nonmagnetic injector is the solid line in
Fig.~\ref{2}(c). Such boundary condition precludes current self-oscillations due to dipole
recycling. Thus, dipole recycling occurs only for the magnetic and successive QWs.

\begin{figure}[ht]
\centerline{\hbox{
\epsfxsize=80mm
\epsfbox{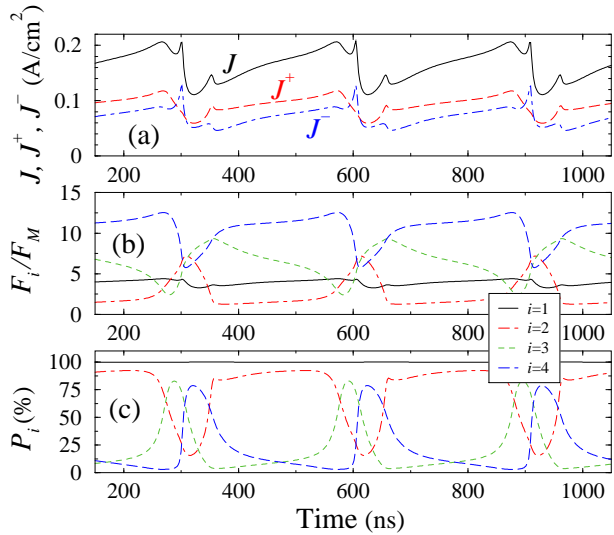}}}
\caption{Tunneling current (a), electric
field (b), and polarization (c) as a function of time, at the i QW
during SSCOs for $N=4$, $V= 0.023$ V, $N_{D}=1.2
\times 10^{10}$cm$^{-2}$ and $F_{M}=0.65$ kV/cm.
Oscillation frequency is 5.4 MHz.}
\label{1}
\end{figure}

Next, we have calculated the shortest SL displaying SSCOs when only the first QW is 
magnetic. For our parameter values, we find SSCOs for SL
having at least 4 periods. Fig.~\ref{1} shows the total current density (most of which is
due to spin-up electrons), the field  and the spin polarization $P_{i}=(n_{i}^+-n_{i}^-)
/(n_{i}^+ + n_{i}^-)$ at the QWs during SSCOs for $N=4$. Note that QW $i=1$
is always fully polarized, whereas the others are strongly polarized only
when the dipole wave is traversing them: their polarizations drop abruptly afterward.
The fraction of the oscillation period during which the $i$th QW is strongly polarized
decreases as $i$ increases.

For $N\geq 4$, SSCOs appear if $N_{D}>N_{D,1}$. We have sought this critical doping 
density for $4\leq N\leq 50$: $N_{D,1} = 2\times 10^{10}/(N-2)\, \mbox{cm}^{-2}$.
In the continuum limit ($N\to \infty$), this approximate formula yields
$N N_{D,1} \approx 2\times 10^{10}$cm$^{-2}$, according to the N-L criterion in
the theory of the Gunn effect \cite{kroemer}.

Our results could be used to construct an oscillatory spin
polarized current injector. A short such device (with 4 QWs)
would inject mostly negatively polarized current whereas long
devices would inject predominantly positively polarized current.
It is important that normal contacts can be used to build the
oscillator, because the crucial requirement is to dope the first
QW with Mn. We have also indicated the range of $N_{D}$
needed to achieve spin polarized SSCOs. For self-oscillations to occur,
appropriate ranges of spin splitting should be induced by tailoring the magnetic
impurity density and external magnetic fields \cite{delta}.

We thank D. S\'anchez for a critical reading of the manuscript and useful suggestions.
This work has been supported by the MECD grants MAT2005-05730-C02-01 and
MAT2005-06444.

\newpage

\noindent FIGURE CAPTIONS
\bigskip
\bigskip

\noindent FIGURE 1. Electric field distribution if the magnetic QW is: (a) $i=1$, (b) $i=20$.
(c) Solid line: $J_{i\to i+1}(F)$ for nonmagnetic $i$ and $i+1$.
For magnetic $i$, nonmagnetic $i\pm 1$: $J_{i\to i+1}$ (dotted line), $J^+_{i\to i+1}$ (dot-dashed line),
$J^-_{i\to i+1}$ (triangles), $J_{i-1\to i}$ (dashed line).
(d) same at larger electric fields.
Parameter values: $N=50$, $V=0.048$ V, $N_{D}=10^{10}$cm$^{-2}$, $F_{M}= 0.64$ kV/cm,
$J_{M}= 0.409$ A/cm$^2$.
\bigskip
\bigskip
\bigskip
\bigskip
\bigskip

\noindent FIGURE 2. Tunneling current (a), electric
field (b), and polarization (c) as a function of time, at the i QW
during SSCOs for $N=4$, $V= 0.023$ V, $N_{D}=1.2
\times 10^{10}$cm$^{-2}$ and $F_{M}=0.65$ kV/cm.
Oscillation frequency is 5.4 MHz.

\end{document}